\documentclass{amsart}

\usepackage{setspace}
\usepackage{graphicx} 
\usepackage{array} 
\usepackage{bm} 
\usepackage{booktabs} 
\usepackage{ctable} 
\usepackage{subfig} 
\usepackage{caption} 
\usepackage{footnote}
\usepackage{threeparttable}
\usepackage{enumerate}
\usepackage{mathtools}
\usepackage{xcolor}
\usepackage{rotating}
\usepackage{lscape}
\usepackage{tensor}
\usepackage[linewidth=1.5pt]{mdframed}
\usepackage{xcolor,colortbl}
\usepackage{pifont} 
\usepackage{hyperref} 

\usepackage[round]{natbib}
\makeatletter 
\renewcommand\@biblabel[1]{#1.} 
\makeatother %

\definecolor{Gray}{gray}{0.85}

\graphicspath{{FiguresDraft/}} 

\numberwithin{equation}{section}

\DeclareMathOperator{\Bern}{Bern}

\DeclareMathOperator{\Exp}{Exp}
\DeclareMathOperator{\Unif}{Unif}
\DeclareMathOperator{\Gam}{Ga}

\DeclareMathOperator{\Dir}{Dir}

\DeclareMathOperator{\PP}{\mathbf{P}}

\DeclareMathOperator{\upi}{\underline\pi}

\begin{document}

\title[Algorithms and diagnostics for the Extended Plackett-Luce model]{Algorithms and diagnostics for the analysis of\\preference rankings with the Extended Plackett-Luce model}

\author
[C.       Mollica]
{Cristina Mollica}
\address
{Dipartimento di Metodi e Modelli per il Territorio, l'Economia e la Finanza\\ 
Sapienza Universit\`a di Roma\\
Via del Castro Laurenziano 9\\00161~Roma\\Italy}
\email{cristina.mollica@uniroma1.it}
\author
[L.   Tardella]
{Luca Tardella}
\address
{Dipartimento di Scienze statistiche\\
 Sapienza Universit\`a di Roma\\
 Piazzale A. Moro 5\\00185~Roma\\Italy}
\email{luca.tardella@uniroma1.it}
\thanks
{Version \today} 



\begin{abstract}

Choice behavior and preferences typically involve numerous and subjective aspects that are difficult to be identified and quantified. For this reason, their exploration is frequently conducted through the collection of ordinal evidence in the form of ranking data.
A \textit{ranking} is an ordered sequence resulting from the comparative evaluation of a given set of \textit{items} according to a specific criterion. Multistage ranking models, including the popular Plackett-Luce distribution (PL), rely on the assumption that the ranking process is performed sequentially, by assigning the positions from the top to the bottom one (\textit{forward order}). A recent contribution to the ranking literature relaxed this assumption with the addition of the discrete \textit{reference order} parameter, yielding the novel \textit{Extended Plackett-Luce model} (EPL). Inference on the EPL and its generalization into a finite mixture framework was originally addressed from the frequentist perspective. 
In this work, we propose the Bayesian estimation of the EPL with order constraints on the reference order parameter. 
The restrictions for the discrete parameter reflect a meaningful rank assignment process and, in combination with the 
data augmentation strategy and the conjugacy of 
the Gamma prior distribution with the EPL,
facilitate the construction
of a tuned joint Metropolis-Hastings algorithm within Gibbs sampling
to simulate from the posterior distribution.
We additionally propose a novel model diagnostic to assess the adequacy of the EPL parametric specification.
The usefulness of the proposal is illustrated with applications to simulated and real datasets.

\end{abstract}
\maketitle
%
\makeatletter \@setaddresses \makeatother \renewcommand{\addresses}{}



\section{Introduction}
\label{s:intro}

Ranking data are common in those experiments aimed at exploring preferences, attitudes or, more generically, choice behavior of a given population towards a set of \textit{items} or \textit{alternatives}~\citep{Vigneau1999,Yu2005,Gormley:Murphy-Royal,Vitelli}. A similar evidence emerges also in the racing context, yielding an ordering of the competitors, for instance players or teams, in terms of their ability or strength, see
~\cite{Henery-Royal},~\cite{Stern1990} and~\cite{Caron:Doucet}.

Formally, a \textit{ranking} $\pi=(\pi(1),\dots,\pi(K))$ of $K$ items is a sequence where the entry $\pi(i)$ indicates the rank attributed to the $i$-th alternative. Data can be equivalently collected in the ordering format $\pi^{-1}=(\pi^{-1}(1),\dots,\pi^{-1}(K))$, such that the generic component $\pi^{-1}(j)$ denotes the item ranked in the $j$-th position. Regardless of the adopted format, ranked observations are multivariate and, specifically,
correspond to permutations of the first $K$ integers. 

The statistical literature concerning ranked data modeling and analysis is reviewed in \cite{Marden} and, more recently, in \cite{Alvo}. Several parametric distributions on the set of permutations $\mathcal{S}_K$ have been developed and applied to real experiments. A popular parametric family is the \textit{Plackett-Luce} model (PL), belonging to the class of the so-called \textit{stagewise ranking models}. The basic idea is the decomposition of the ranking process into $K-1$ stages, concerning the attribution of each position according to the \textit{forward order}, that is, the ordering of the alternatives proceeds sequentially from the most-liked to the least-liked item. The implicit forward order assumption has been released by~\cite{Mollica:Tardella} in the \textit{Extended Plackett-Luce model} (EPL). The PL extension relies on the introduction of the \textit{reference order} parameter, indicating the rank assignment order, and its estimation was originally considered from the frequentist perspective.

In this work, we 
investigate
a restricted version of the EPL with order constraints for the reference order parameter and describe an original MCMC method to perform Bayesian inference. In particular, the considered parameter constraints formalize a meaningful
rank attribution process and we show how they can facilitate the definition of a joint proposal distribution for the Metropolis-Hastings (MH) step. Moreover,
we introduce
a novel diagnostic to assess the adequacy of the EPL assumption as the actual sampling distribution of the observed rankings.

The outline of the article is the following. 
After a review of
the main features of the EPL and the related data augmentation approach with latent variables,
the novel Bayesian EPL with order constraints is introduced in Section~\ref{s:pl}.
The detailed description of the MCMC algorithm to perform approximate posterior inference is presented in Section~\ref{s:MCMC}, whereas a novel diagnostic for the EPL specification is 
argued
in Section~\ref{s:moddiag} with
illustrative applications
to both simulated and
real ranking data.


\section{The Bayesian Extended Plackett-Luce model}
\label{s:pl}


\subsection{Model specification}
\label{ss:modspec}
%

\begin{figure}
\begin{center}
\includegraphics[height=6cm, width=6cm]{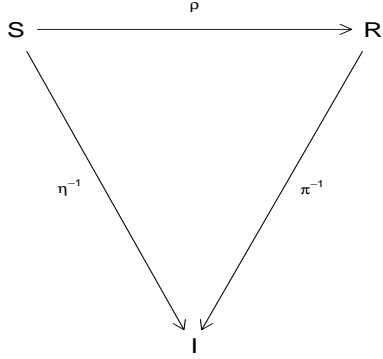}
\caption
{Mappings between the set of items, ranks and stages.}
\label{fig:mappings}
\end{center}
\end{figure}

The PL
was introduced by~\cite{Luce} and~\cite{Plackett} and has a long history in the ranking literature for its numerous successful applications as well as for still inspiring new research developments.
The PL is a parametric class of ranking distributions indexed by the
\textit{support parameters} $\underline{p}=(p_1,\dots,p_K)$,
representing positive measures of liking for each item:
the higher the value of the support parameter $p_i$,
the greater the probability for the $i$-th item to be preferred at each selection stage. 
The expression of the PL distribution is
\begin{equation}
\label{e:pl}
\PP_\text{PL}(\pi^{-1}|\underline{p})
=\prod_{t=1}^K\frac{p_{\pi^{-1}(t)}}{\sum_{v=t}^Kp_{\pi^{-1}(v)}}\qquad\pi^{-1}\in\mathcal{S}_K,
\end{equation}
revealing the analogy of the underling ranking 
selection
process with the sampling without replacement of the alternatives in order of preference. 
The implicit assumption in the PL scheme is the forward ranking order, meaning that at the first stage the ranker reveals the item in the first position (most-liked alternative), at the second stage she assigns the second position and so on up to the last rank (least-liked alternative). \cite{Mollica:Tardella} suggested the extension of the PL by relaxing the canonical forward order assumption, in order to explore alternative meaningful ranking orders for the choice process and to increase the flexibility of the PL parametric family. Their proposal was realized by representing the ranking order with an additional
model parameter  $\rho=(\rho(1),\dotsc,\rho(K))$, 
called reference order and defined as the bijection between the stage set $S=\{1,\dots,K\}$ and the rank set $R=\{1,\dots,K\}$
\begin{equation*}
\rho:S\to R,
\end{equation*}
such that the entry $\rho(t)$ indicates the rank attributed at the $t$-th stage of the ranking process. 
Thus, $\rho$ is
a
discrete 
parameter given by a permutation of the first $K$ integers
and the composition $\eta^{-1}=\pi^{-1}\rho$ of an ordering
with a reference order 
yields the bijection between the stage set $S$ and the item set $I=\{1,\dots,K\}$
\begin{equation*}
\label{e:eta}
\eta^{-1}:S\to I.
\end{equation*}
The sequence $\eta^{-1}=(\eta^{-1}(1),\dotsc,\eta^{-1}(K))$ lists the items in order of selection, such that the component $\eta^{-1}(t)=\pi^{-1}(\rho(t))$ corresponds to the item chosen at stage $t$ and receiving rank $\rho(t)$. Figure~\ref{fig:mappings} summarizes all the possible sequences defined as bijective mappings between the set of items, ranks and stages. 

The probability of a generic ordering under EPL can be written as
\begin{equation}
\label{e:EPL}
\PP_\text{EPL}(\pi^{-1}|\rho,\underline{p})=\PP_\text{PL}(\pi^{-1}\rho|\underline{p})
=\prod_{t=1}^K\frac{p_{\pi^{-1}(\rho(t))}}{\sum_{v=t}^Kp_{\pi^{-1}(\rho(v))}}\qquad\pi^{-1}\in\mathcal{S}_K,
\end{equation}
%
%
Hereinafter, we will shortly refer to~\eqref{e:EPL} as $\text{EPL}(\rho,\underline{p})$. The quantities $p_i$'s
are still proportional to the probabilities for each item to be selected at the first stage,
but to be ranked
in the position
indicated by the first entry of $\rho$. Obviously, the standard PL is the special instance of the EPL with the forward order $\rho_\text{F}=(1,2,\dots,K)$, known as the \textit{identity permutation}. Similarly, with the backward order $\rho_\text{B}=(K,K-1,\dots,1)=(K+1)-\rho_\text{F}$, one has the backward PL.

As in \cite{Mollica:Tardella2017}, the data  augmentation with
the latent quantitative variables $\underline{y}=(y_{st})$ for $s=1,\dots,N$ and $t=1,\dots,K$ crucially contributes to make the Bayesian inference for the EPL analytically tractable. The complete-data model can be specified as follows
%
%
\begin{eqnarray*}
\pi_s^{-1}|\rho,\underline{p} & \overset{\text{iid}}{\sim} & \text{EPL}(\rho,\underline{p})\qquad\qquad\quad\qquad s=1,\dots,N,\\
y_{st}|\pi_s^{-1},\rho,\underline{p} & \overset{\text{i}}{\sim} & \Exp\left(\sum_{\nu=t}^{K}p_{\pi_s^{-1}(\rho(\nu))}\right)\qquad t=1,\dots,K,
\end{eqnarray*}
where the auxiliary variables $y_{st}$'s are assumed to be conditionally independent on each other and exponentially distributed with rate parameter equal to the normalization term of the EPL. The complete-data likelihood turns out to be
%
\begin{equation}
\label{e:complik}
L_c(\rho,\underline{p},\underline{y})=\prod_{i=1}^Kp_i^Ne^{-p_i\sum_{s=1}^N\sum_{t=1}^Ky_{st}\delta_{sti}}
\end{equation}
where
%
\begin{equation*}
\delta_{sti}=\begin{cases}
      1\quad\text{ if }i\in\{\pi_s^{-1}(\rho(t)),\dots,\pi_s^{-1}(\rho(K))\},\\
      0\quad\text{ otherwise},
\end{cases}
\end{equation*}
with $\delta_{s1i}=1$ for all $s=1,\dots,N$ and $i=1,\dots,K$.


\subsection{Parameter space and prior distributions}
\label{ss:prior}
For the prior specification, we consider independence of $\underline{p}$ and $\rho$ and the following distributions
\begin{eqnarray*}
p_i & \overset{\text{i}}{\sim} & \Gam(c,d)\qquad i=1,\dots,K,\\
\rho & \sim & \Unif\left\{\tilde{\mathcal{S}}_K\right\}.
\end{eqnarray*}
The adoption of independent Gamma densities for the support parameters is motivated by the conjugacy with the model, as apparent by checking the form of the likelihood \eqref{e:complik}.
Differently from \cite{Mollica:Tardella}, we 
focus on a restriction $\tilde{\mathcal{S}}_K$ of the whole permutation space $S_K$ for the generation of the reference order. Our choice can be explained by the fact that, in a preference elicitation process, not all the possible $K!$ orders seem to be equally natural, hence plausible.
Often the ranker has a clearer perception about her extreme preferences (most-liked and least-liked items), rather than middle positions. In this perspective,
the rank attribution process can be regarded as the result of a sequential ``top-or-bottom'' selection of the positions. At each stage, the ranker specifies either her best or worst choice among the available positions at that given step.
\begin{figure}
\begin{center}
\includegraphics[height=5cm, width=10cm]{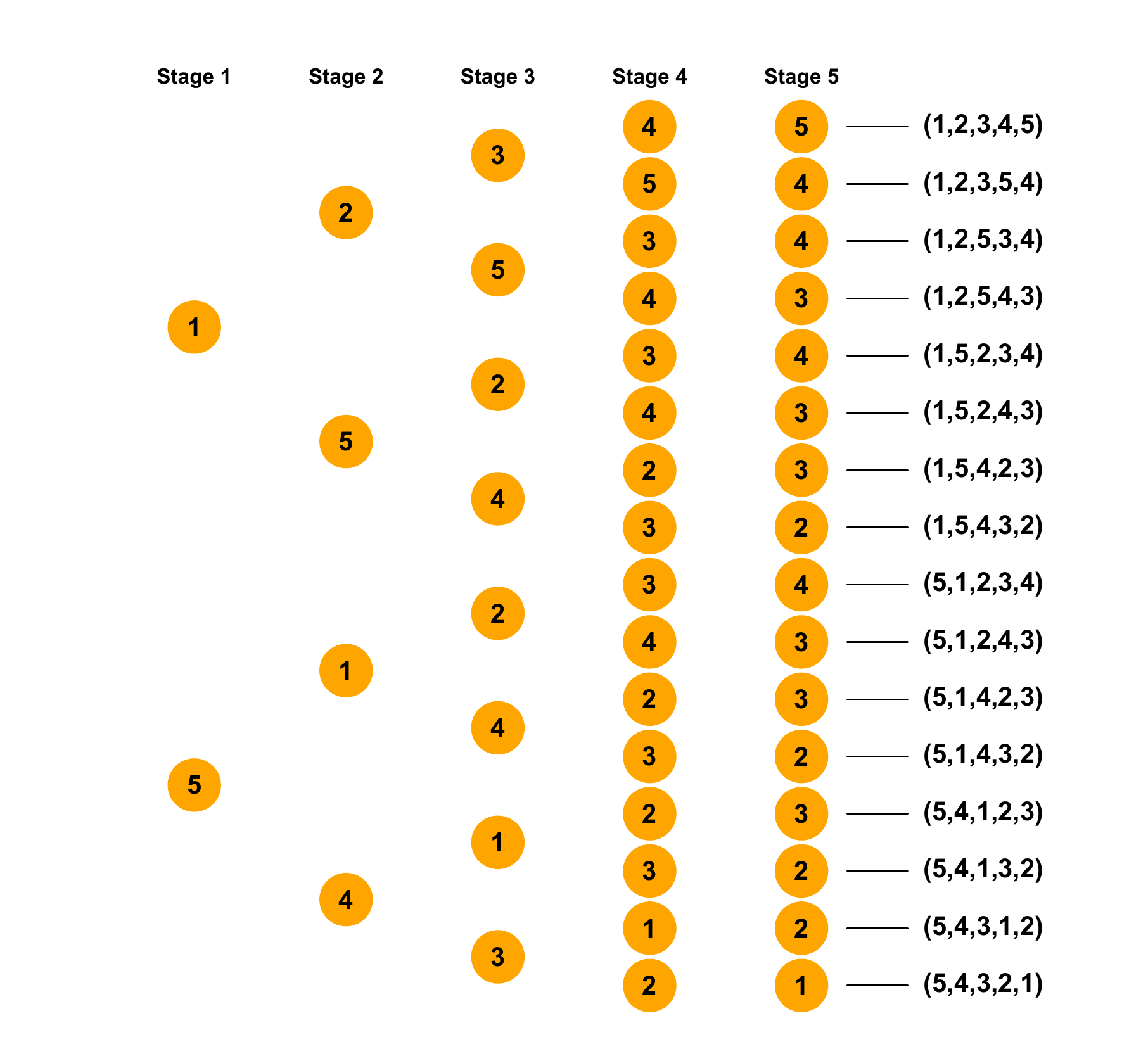}
\caption
{Restricted permutation space $\tilde S_5$ for the reference order parameter $\rho$ in the case of $K=5$ items.}
\label{fig:permtree}
\end{center}
\end{figure}
%
Figure~\ref{fig:permtree} shows the restricted permutation space $\tilde S_5$ for the reference order in the case of $K=5$ items. The first entry of the reference order, indicating the position assigned at the first stage, can be either $\rho(1)=1$ (most-liked item) or $\rho(1)=5$ (least-liked item). Let us suppose that 
at the first stage the ranker has ranked the item in the last position i.e. $\rho(1)=5$; 
at the second stage, the ranker can express only her best or worst choice, conditionally on the fact that the last position has been already occupied; this means that either $\rho(2)=1$ or $\rho(2)=4$, and so on up to the final stage where the last component $\rho(K)$ is automatically determined. 
With this scheme, the reference order can be equivalently represented as a binary sequence $\underline{W}=(W_1,\dots,W_K)$
where the generic $W_t$ component indicates whether the ranker makes a top or bottom decision at the $t$-th stage,
with the convention that $W_K=1$.
One can 
then formalize the mapping from the restricted permutation $\rho$ to  $\underline{W}$ 
with the help of a vector of non negative integers $\underline{F}=(F_1,\dots,F_K)$, 
where $F_t$ represents the number of top positions assigned before stage $t$.
In fact, by starting from positing by construction $F_1=0$, one can derive sequentially
\begin{equation*}
W_t=I_{[\rho(t)=\rho_\text{F}(F_t+1)]}
=\begin{cases}
      1\qquad \text{at stage $t$ the top preference is specified}, \\
      0\qquad \text{at stage $t$ the bottom preference is specified},
\end{cases}
\end{equation*}
where $I_{[E]}$ is the indicator function of the event $E$ and  
$F_t=\sum_{\nu=1}^{t-1}W_\nu$ for $t=2,...,K$.
Note that, since the forward and backward orders $(\rho_\text{F}, \rho_\text{B})$ 
can be regarded as the two extreme benchmarks in the sequential construction of $\rho$,
this allows us to understand that 
$\rho_F(F_t+1)$ corresponds to the top 
position available at stage $t$.
Conversely, $B_t=(t-1)-F_t$
is the number of bottom positions assigned before stage $t$ and thus,
symmetrically, one can understand that $\rho_B(B_t+1)$ indicates the bottom
position available at stage $t$.

In order to clarify the restriction on the reference order space and the 
related notation, let us make an example. The reference order $\rho=(5,1,4,3,2)$ can be coded in binary format as follows
\begin{align*}
F_1&=0,\qquad B_1=0,\qquad W_1=I_{[\rho(1)=\rho_\text{F}(1)]}=I_{[5=1]}=0,\\
F_2&=0,\qquad B_2=1,\qquad W_2=I_{[\rho(2)=\rho_\text{F}(1)]}=I_{[1=1]}=1,\\
F_3&=1,\qquad B_1=1,\qquad W_1=I_{[\rho(3)=\rho_\text{F}(2)]}=I_{[4=2]}=0,\\
F_4&=1,\qquad B_1=2,\qquad W_1=I_{[\rho(4)=\rho_\text{F}(2)]}=I_{[3=2]}=0,\\
F_5&=1,\qquad B_5=3,\qquad W_5=I_{[\rho(5)=\rho_\text{F}(2)]}=I_{[2=2]}=1,\\
\end{align*}
implying $\underline{F}=(0,0,1,1,1)$, $\underline{B}=(0,1,1,2,3)$ and $\underline{W}=(0,1,0,0,1)$. This means that, apart from the second stage, the ranker always specified her preferences by assigning bottom positions. 

The inverse relation from the binary vector $\underline{W}$ to the
constrained reference order $\rho \in \tilde{\mathcal{S}}_K$
can be written as follows
%
\begin{equation*}
\rho(t)=\rho_\text{F}(F_t+1)^{W_t}\rho_\text{B}(B_t+1)^{1-W_t}\qquad t=1,\dots,K.
\end{equation*}
%
So, for the previous example, the inverse mapping turns out to be
%
\begin{align*}
\rho(1) & =\rho_\text{F}(1)^{0}\rho_\text{B}(1)^{1}=\rho_\text{B}(1)=5,\\
\rho(2) & =\rho_\text{F}(1)^{1}\rho_\text{B}(2)^{0}=\rho_\text{F}(1)=1,\\
\rho(3) & =\rho_\text{F}(2)^{0}\rho_\text{B}(2)^{1}=\rho_\text{B}(2)=4,\\
\rho(4) & =\rho_\text{F}(2)^{0}\rho_\text{B}(3)^{1}=\rho_\text{B}(3)=3,\\
\rho(5) & =\rho_\text{F}(2)^{1}\rho_\text{B}(4)^{0}=\rho_\text{F}(2)=2,
\end{align*}
and hence $\rho=(5,1,4,3,2)$.

The binary representation of the reference order suggests that, under the constraints of the ``top-or-bottom'' scheme, the size of $\tilde{\mathcal{S}}_K$ is equal to $2^{K-1}$. The reduction of the reference order space into a finite set with an exponential size, rather than with a factorial cardinality, is convenient for at least two reasons: i) it leads to a more intuitive interpretation of the support parameters, since they become proportional to the probability for each item to be ranked either in the first or in the last position and ii) it facilitates the construction of a MH step to sample the reference order parameter.

\section{Bayesian estimation of the EPL via MCMC}
\label{s:MCMC}
In this section, we describe an original MCMC algorithm to solve the Bayesian inference for the constrained EPL. Specifically, our simulation-based method to approximate the posterior distribution is a tuned joint Metropolis-within-Gibbs sampling (TJM-within-GS), where the simulation of the reference order is accomplished with a MH algorithm relying on a joint proposal distribution on $\rho$ and $\underline{p}$, whereas the posterior drawings of the latent variables $y$'s and the support parameters are performed from the related full-conditional distributions.

\subsection{Tuned Joint Metropolis-Hastings step and Swap move}
\label{ss:tjmh}
Let us denote with $\underline{\lambda}=(\lambda_1,\dots,\lambda_K)$ the vector of Bernoulli probabilities 
$$\lambda_t=\PP(W_t=1|W_1,\dots,W_{t-1})=\PP(\rho(t)=\rho_\text{F}(F_t+1)|\rho(1),\dots,\rho(t-1)).$$ 
A possible proposal distribution to be employed in the MH step for sampling the reference order
could have the following form
\begin{equation*}
\PP(\rho)=\prod_{t=1}^K\PP(\rho(t)|\rho(1),\dots,\rho(t-1))=\prod_{t=1}^K\lambda_t^{W_t}(1-\lambda_t)^{1-W_t}.
\end{equation*}
However, preliminary implementations of a MH-within GS algorithm on synthetic data 
suggested that only a joint proposal distribution of the support parameters and the reference order allows for an adequate mixing of the resulting Markov Chain.
Thus, to simultaneously sample candidate values for $\rho$ and $\underline{p}$, we 
devised the following 
joint proposal distribution $g(\rho,\underline{p})$ with a specific decomposition of the dependence structure, given by
\begin{equation}
\label{e:jp}
g(\rho,\underline{p})=g(\rho(1))\times g(\underline{p}|\rho(1))\times \prod_{t=2}^Kg(\rho(t)|\underline{p},\rho(1),\dots,\rho(t-1)).
\end{equation}
The dependence structure in~\eqref{e:jp} shows that, after drawing the first component of $\rho$, the proposal 
can exploit
the sample evidence on the support parameters to guide the simulation of the remaining candidate entries of the reference order. In so doing, the generation of the two parameter vectors are linked to each other,
in order to mimic the target density and, hence, getting a better mixing chain.
Candidate values $(\tilde\rho,\tilde{\underline{p}})$
are jointly generated according to the following scheme:
\begin{enumerate}
\item sample the first component of $\rho$ (stage $t=1$)
$$\tilde W_1\sim\Bern(\tilde\lambda_1)\quad\Rightarrow\quad\tilde \rho(1)=\rho_\text{F}(1)^{\tilde W_1}\rho_\text{B}(1)^{1-\tilde W_1}=1^{\tilde W_1}K^{1-\tilde W_1}.$$
In our application, we set $\tilde \lambda_1=\PP(\tilde W_1=1)=0.5$;
\item sample the support parameters
$$\tilde{\underline{p}}|\tilde \rho(1)\sim\Dir(\underline{\alpha}_0\times\underline{r}_{\tilde \rho(1)}),$$
where $\underline{\alpha}_0$ is a vector of tuning parameters and $\underline{r}_{\tilde \rho(1)}$ is the vector collecting either the marginal top or bottom item frequencies
according to whether $\tilde\rho(1)=1$ or $\tilde\rho(1)=K$. 
Specifically, the $i$-th entry of $\underline{r}_{\tilde \rho(1)}$ is
$$r_{\tilde \rho(1)i}=\sum_{s=1}^NI_{[\pi_s^{-1}(\tilde \rho(1))=i]}$$
\item to sample the remaining entries of the reference order (from stage $t=2$ to stage $K-1$), we will proceed iteratively as follows: once we have selected the reference order component at stage $t-1$, we consider the two 
observed
contingency tables having as first margin the item 
placed at the current reference order component $\tilde \rho(t-1)$ and as second margin, respectively, the item 
placed at the reference order component which can be possibly selected at the next stage, denoted as either $\rho_F(F_t+1)$
or $\rho_B(B_t+1)$.
The generic entries of the two contingency tables are
%
\begin{align*}
\tilde\tau_{ii't}&=\sum_{s=1}^NI_{[\pi_s^{-1}(\tilde \rho(t-1))=i,\pi_s^{-1}(\rho_F(F_t+1))=i']},\\
\tilde\beta_{ii't}&=\sum_{s=1}^NI_{[\pi_s^{-1}(\tilde \rho(t-1))=i,\pi_s^{-1}(\rho_B(B_t+1))=i']},
\end{align*}
%
%
corresponding to the actually observed joint frequencies counting how many times each item $i$ in the previous stage is followed by any other item $i'$ at the next stage. To compare these frequencies with the corresponding expected frequencies $\tilde E_{ii't}$ under the EPL,
we use a Monte Carlo approximation
$$\tilde\eta_s^{-1}(1),\dots,\tilde\eta_s^{-1}(t)|\tilde{\underline{p}}\overset{\text{i}}{\sim}\text{PL}(\tilde{\underline{p}})\qquad s=1,\dots,N,$$
$$\tilde E_{ii't}=\sum_{s=1}^NI_{[\tilde\eta_s^{-1}(t-1)=i,\tilde\eta_s^{-1}(t)=i']}$$
and compute the following top and bottom distances
$$\tilde d_t^{T}=\sum_{i=1}^K\sum_{i'=1}^K(\tilde\tau_{ii't}-\tilde E_{ii't})^2\quad\text{and}\quad\tilde d_t^{B}=\sum_{i=1}^K\sum_{i'=1}^K(\tilde\beta_{ii't}-\tilde E_{ii't})^2.$$
The above distances are then suitably scaled
as follows
$$\tilde d_{t}=1-\frac{\tilde d_{t}^{T}}{\tilde d_{t}^{T}+\tilde d_{t}^{B}}$$
and exploited in order to mimic the conditional probability corresponding to the target distribution.
Indeed, we define the Bernoulli proposal probability of top selection at stage $t$ as
$$\tilde\lambda_{t}=\tilde d_{t}(1-2h)+h,$$
where $h\in(0,0.5)$ is a tuning parameter introduced to guarantee
a minimal positive probability $h$ for the bottom selection ($\tilde\lambda_{t} \geq h$). We set as default value $h=0.1$. Finally, for $t=2,\dots,K$, we sample
$$\tilde W_t\sim\Bern(\tilde\lambda_t)\quad\Rightarrow\quad\tilde \rho(t)=\rho_F(F_t+1)^{\tilde W_t}\rho_B(B_t+1)^{1-\tilde W_t}.$$
%
\end{enumerate}
The 
resulting
joint proposal probability of the candidate values
is
$$g(\tilde\rho,\tilde{\underline{p}})=\Dir(\tilde{\underline{p}}|\underline{\alpha}_0\times\underline{r}_{\tilde \rho(1)})\prod_{t=1}^K\tilde\lambda_{t}^{\tilde W_t}(1-\tilde\lambda_{t})^{(1-\tilde W_t)}.$$
Hence, 
if we denote with $L(\rho,\underline{p})$ the observed-data likelihood, the acceptance probability turns out to be
\begin{equation*}
\alpha'=
\min\left\{\frac{g(\rho^{(l)},\underline{p}^{(l)})}{g(\tilde\rho,\tilde{\underline{p}})}
\frac{L(\tilde\rho,\tilde{\underline{p}})\prod_{i=1}^K\Gam(\tilde p_i|c,d)}{L(\rho^{(l)},\underline{p}^{(l)})\prod_{i=1}^K\Gam(p_i^{(l)}|c,d)},1\right\}.
\end{equation*}
The MH step ends with the classical acceptance/rejection of the candidate pair
%
\begin{equation*}
(\rho',\underline{p}')=\begin{cases}
      (\tilde\rho,\tilde{\underline{p}})\qquad \text{\qquad if $\log(u')<\log(\alpha')$}, \\
      (\rho^{(l)},\underline{p}^{(l)})\qquad \text{otherwise},
\end{cases}
\end{equation*}
where $u'\sim\Unif(0,1)$. 

In this phase of the algorithm, the values $(\rho',\underline{p}')$ have to be regarded as temporary parameter drawings. 
In fact, we use a composition of three kernels, where the intermediate $(\rho',\underline{p}')$ comes from the first MH kernel just described 
and is such that 
the possibly accepted $\underline{p}'$ 
should improve the sampling of the discrete parameter. 
Indeed, the second kernel is just a full Gibbs sampling cycle involving the $(\underline{p},\underline{y})$ components, detailed in Section~\ref{ss:tjmhgibbs}. 

\subsection{Swap move}
\label{ss:tjmh}
We now describe how to complete the joint update of the $(\rho,\underline{p})$ parameter vectors.
In order to further accelerate the exploration of the parameter space, we propose for the $\rho$ component only a third MH step (kernel), that accomodates for a possible local move w.r.t. to the current value $\rho'$. The idea relies on a random swap of two adjacent components of $\rho'$, that we refer to as \textit{Swap Move} (SM). Let $M\in\{1,\dots,K-1\}$ be the number of applicable contiguous swaps on $\rho'$, such that the order constraints of $\tilde{\mathcal{S}}_K$ still hold in the returning sequence, and $\{t_1,\dots,t_M\}$ be the indexes of the entries of $\rho'$ that can be switched with the consecutive ones. Note that the last two entries can be always swapped, meaning that $t_M=K-1$. The additional MH step consists in proposing a further reference order with a randomly selected SM. Specifically, one first simulate 
$$t^*\sim\Unif\{t_1,\dots,t_M\}$$
and then define the new candidate as
$$\rho''=(\rho'(1),\dots,\rho'(t^*+1),\rho'(t^*),\dots,,\rho'(K)).$$
Finally, by computing the acceptance probability as
\begin{equation*}
\alpha''=
\min\left\{\frac{g(\rho',\underline{p}')}{g(\rho'',\underline{p}')}
\frac{L(\rho'',\underline{p}')}
{L(\rho',\underline{p}')}
,1\right\},
\end{equation*}
the sampled value of the reference order at the $(l+1)$-th iteration turns out to be
\begin{equation*}
\rho^{(l+1)}=\begin{cases}
      \rho''\qquad \text{if $\log(u'')<\log(\alpha'')$}, \\
      \rho'\qquad \text{otherwise},
\end{cases}
\end{equation*}
where $u''\sim\Unif(0,1)$.

\subsection{Tuned Joint Metropolis-within-Gibbs-sampling}
\label{ss:tjmhgibbs}
At the generic iteration $l+1$, the TJM-within-GS
iteratively alternates the following simulation steps
\begin{eqnarray*}
\rho^{(l+1)},\underline{p}' & \sim & \text{TJM + SM},\\
y_{st}^{(l+1)}|\pi_s^{-1},\rho^{(l+1)},\underline{p}' & \sim & \Exp\left(\sum_{i=1}^K\delta_{sti}^{(l+1)}p_i'\right),\\
p_{i}^{(l+1)}|\underline\pi^{-1},\underline{y}^{(l+1)},\rho^{(l+1)} & \sim & \text{Ga}\left(c+N,d+\sum_{s=1}^{N}\sum_{t=1}^{K}\delta_{sti}^{(l+1)}y_{st}^{(l+1)}\right).
\end{eqnarray*}
The above outline shows that the full-conditional of the unobserved continuous variables $y$'s is given by construction of the complete-data model specified in Section~\ref{ss:modspec}, whereas the full-conditional of the support parameters is induced by the conjugate structure, requiring a straightforward update of the corresponding Gamma prior.

\section{Model diagnostic}
\label{s:moddiag}

Simulation studies confirmed the efficacy of the TJM-within-GS to recover the actual generating EPL, together with the benefits of the SM strategy to speed up the MCMC algorithm in the exploration of the posterior distribution. However, 
we were surprised to verify a less satisfactory performance of the TJM-within-GS in terms of posterior exploration in the application to some real-world examples, such as the famous \texttt{song} dataset analyzed by \cite{Fligner:Verducci-MathPsycho}.
Since the joint proposal distribution relies on summary statistics, the posterior sampling procedure is expected to work well as long as the data are actually taken from an EPL distribution. So, the 
unexpectedly bad 
behavior of the MCMC algorithm suggested 
to conjecture that,
for such real data, the EPL does not represent the true (or in any case an appropriate) data generating mechanism.
This has motivated us  to the develop
some new
tools to appropriately 
check the model mis-specification issue. Indeed, few proposals in the ranking literature concern the construction of diagnostics for model adequacy. 
The frequentist ones have been somehow reviewed in \cite{Mollica:Tardella2017} and have been used to derive some appropriate posterior predictive checks.
In the next section, we 
propose
an original diagnostic for the EPL specification.

\subsection{EPL diagnostic}
\label{ss:modest}

Suppose we have some data simulated from an EPL model. We expect the marginal frequencies of the items at the first stage to be ranked according to the order of the corresponding support parameter component. On the other hand, 
although difficult to derive exactly, we expect the marginal frequencies of the items at the last stage to be ranked according to the reverse order of the corresponding support parameter component. If one can prove such a statement, one can then derive that the ranking of the marginal frequencies of the items corresponding to the first and last stage should sum up to $(K+1)$, no matter what their support is. Of course, this is less likely to happen when the sample size is small or when the support parameters are not so different of each other. In any case, one can define a test statistic by considering,
for each couple of integers $(j,j')$ candidate to represent the first and the last stage ranks, namely $\rho(1)$ and $\rho(K)$, 
a discrepancy measure $T_{jj'}(\upi)$
between $K+1$ and the
sum of the rankings of the
frequencies corresponding to the same item extracted in the first and in the last stage. Formally, let $\underline{r}^{[1]}_j=(r^{[1]}_{j1},\dots,r^{[1]}_{jK})$ and 
$\underline{r}^{[K]}_{j'}=(r^{[K]}_{j'1},\dots,r^{[K]}_{j'K})$ be
the marginal item frequency distributions for the $j$-th  and $j'$-th positions, 
to be assigned respectively at the first [1] and last [K] stage.
In other words, the generic entry $r^{[s]}_{ji}$ is the number of times that item
$i$ is ranked $j$-th at the $s$-th stage. The proposed EPL diagnostic relies on the following discrepancy
\begin{equation*}
T_{jj'}(\upi)=\sum_{i=1}^K\lvert(\text{rank}(\underline{r}^{[1]}_{j})_i+\text{rank}(\underline{r}^{[K]}_{j'})_i-(K+1))\rvert,
\end{equation*}
implying that the 
smaller the test statistics, the larger the plausibility that the two integers $(j,j')$ represent the first and the last components of the reference order. 
This tends to happen 
more often for larger sample sizes. To globally assess the conformity of the sample with the EPL, we consider the minimum value of $T_{jj'}(\upi)$ over all the possible rank pairs satisfying the order constraints
%
\begin{equation}
\label{t:T}
T(\upi)=\underset{(j,j')\in\mathcal P}{\text{min}}\,T_{jj'}(\upi),
\end{equation}
where $\mathcal P=\{(j,j'): j\in \{1,K\}\text{ and }j\neq j'\}$.
Note that the diagnostic~\eqref{t:T} can be generalized by considering all the possible rank pairs $\{(j,j'): j\neq j'\}$ to assess the plausibility of an unrestricted (without order constraints) EPL. 

\section{Illustrative applications}
\label{s:appl}

\subsection{Applications to simulated data}
\label{ss:appsim}
To check the efficacy of our proposal, we started by simulating $N=100$ orderings of $K=5$ items from a genuine EPL model with a parameter configuration given by
$$\rho=(1,5,2,4,3)\quad\text\quad\underline{p}=(0.15,0.4,0.12,0.08,0.25).$$
Under the above EPL specification, the expected rankings of the items in order of occurrence at the first and the last stage are indicated in Table~\ref{t:firstlast}.
\begin{table}[t]
\caption{Expected rankings of the items in terms of number of selections at the first and the last stage for a simulated sample from $\text{EPL}(\rho=(1,5,2,4,3),\underline{p}=(0.15,0.4,0.12,0.08,0.25))$. The true first and last stage ranks correspond, respectively, to rank 1 and 3.}
\label{t:firstlast}
\centering
\begin{tabular}{cccccc}
 & \multicolumn{5}{c}{Item} \\ 
\cline{2-6}  
 & 1 & 2 & 3 & 4 & 5 \\ 
\cline{2-6}  
$\text{rank}(\underline{r}^{[1]}_{1})$ & 3 & 1 & 4 & 5 & 2  \\
$\text{rank}(\underline{r}^{[K]}_{3})$  & 3 & 5 & 2 & 1 & 4  \\
\hline
Sum of ranks  & 6 & 6 & 6 & 6 & 6  \\
\end{tabular}
\end{table}
The matrix with the entries $T_{jj'}(\upi)$ of the diagnostic for all pairs $j\neq j'$ is shown in Table~\ref{t:Tmatrix}. 
\begin{table}[t]
\caption{Entries of the EPL diagnostic for a simulated sample from $\text{EPL}(\rho=(1,5,2,4,3),\underline{p}=(0.15,0.4,0.12,0.08,0.25))$. The true first and last stage ranks correspond respectively to rank 1 and 3, yielding the minimum value of the diagnostic (in bold).}
\label{t:Tmatrix}
\centering
\begin{tabular}{cccccc}
 & \multicolumn{5}{c}{$j'$} \\ 
\cline{2-6}  
$j$ & 1 & 2 & \bf{3} & 4 & 5 \\ 
\hline
\rowcolor{Gray}
\bf{1} & - & 12 & \bf{1} & 6 & 10  \\
2  & 12 & - & 2 & 4 & 11  \\
3 & 1 & 2 & - & 9 & 5  \\
4  & 6 & 4 & 9 & - & 10  \\
\rowcolor{Gray}
5 & 10 & 11 & 5 & 10 & -  \\
\hline
\end{tabular}
\end{table}
For the constrained EPL, we have to focus on the first and last row of such a matrix (highlighted in grey), yielding the observed value $T(\upi)=1$ of the test statistic for the rank pair $(j,j')=(1,3)$, which actually is the global minimum of the whole matrix as well as the pair of the true first and last stage ranks. 
Now one 
can check that it is not unlikely to have a value of the theoretical reference distribution under the assumed model which is greater than or equal to the observed test statistic, for example with a $p$-value computed with the bootstrap (frequentist) method or via posterior predictive (Bayesian) simulation. Deviations from the EPL model should yield greater values of the test statistic and hence smaller $p$-values. By applying the bootstrap approach with 100 datasets drawn from the observed sample, we obtained $p$-value = 0.48, which is in line with the value 0.50 expected under a correct model specification.

We further extended the simulation study to enquire into the power function of the novel test statistic $T(\upi)$. We drew 50 samples composed of $N=149$ orderings of $K=5$ items from
\begin{itemize}
\item[-] an EPL($\rho=(1,5,2,4,3), \underline{p}=(0.15,0.4,0.12,0.08,0.25)$), to assess the rate of Type I error;
\item[-] a Mallows model with modal ranking $\sigma=(1,2,3,4,5)$ and Hamming distance $ d_H= $ 2, see \cite{Diaconis:LectureNotes}, to assess the power of the test ($1-\text{rate of Type II error}$).
\end{itemize}
Under the two population scenarios, we estimated the probability of rejecting the null hypothesis that the data were generated from an EPL with the relative frequency of the times that the bootstrap $p$-values are smaller than or equal to the typical critical threshold $\alpha=0.05$. In particular, from our simulation study we estimated a rate of incorrectly rejecting the null hypothesis equal to 0, whereas the rate of rejection under the Mallows model was estimated equal to 0.6.

\subsection{Applications to real data}
\label{ss:appsim}

\begin{figure}
\begin{center}
\includegraphics[width=12cm,height=8cm]{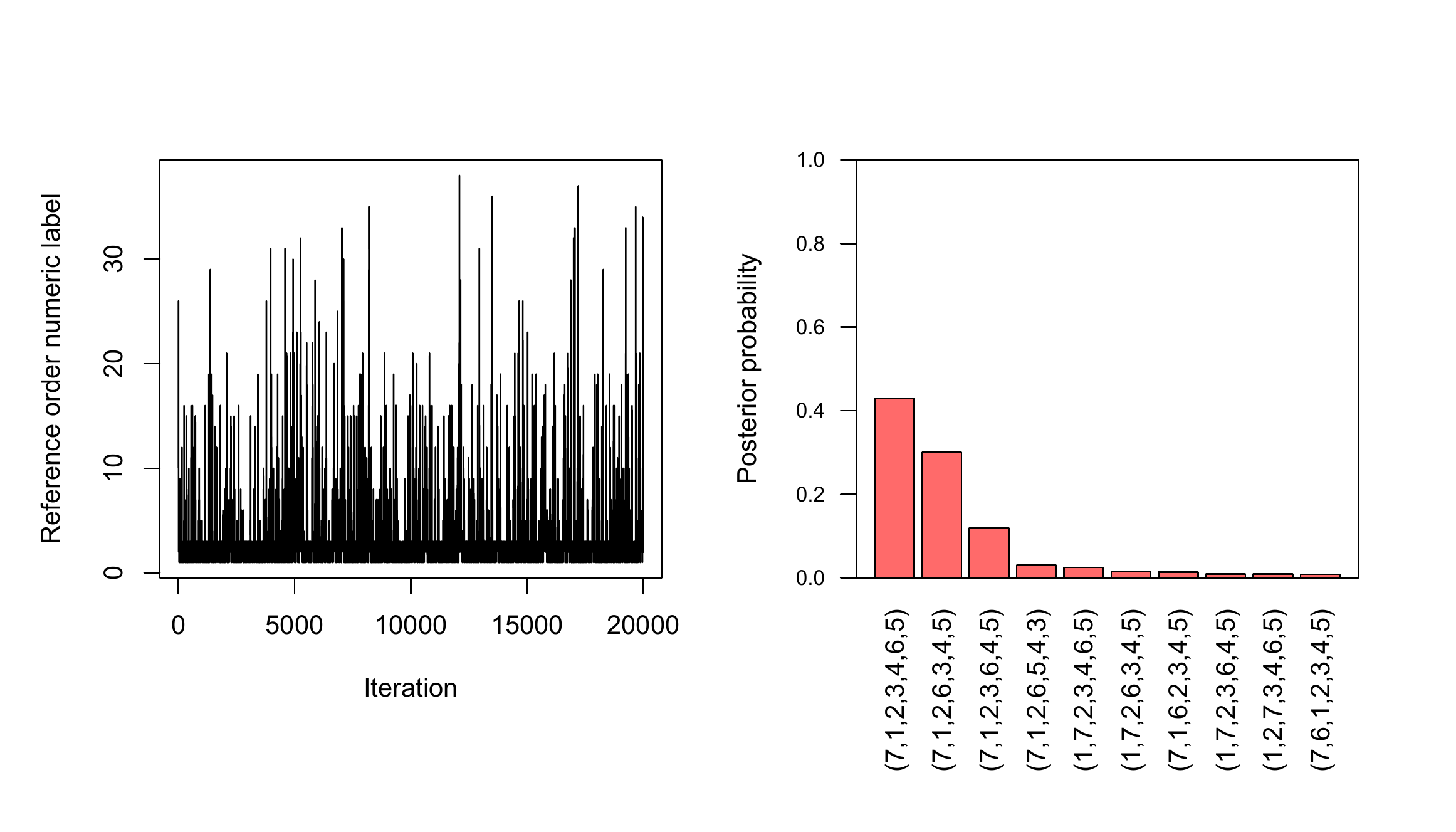}%
\caption{Traceplot (left) and top-10 posterior probabilities (right) for the reference order parameter.}
\label{fig:postref}
\end{center}
\end{figure}

We fit the EPL with reference order constraints to the \texttt{sport}
dataset of the \texttt{Rankcluster} package, where $N$=130 students at
the University of Illinois were asked to rank $K$=7 sports in order of
preference:  $1$=Baseball, $2$=Football, $3$=Basketball, $4$=Tennis, $5$=Cycling, $6$=Swimming and 
$7$=Jogging. We estimated the Bayesian EPL with hyperparameter setting $c=d=1$, by running the TJM-within-GS described in Section~\ref{s:MCMC} for 20000 iterations and discarding the first 2000 samplings as burn-in phase. 
We show
the
approximation of the posterior distribution on the reference
order in Figure~\ref{fig:postref}, where it is apparent that the MCMC is mixing sufficiently fast
and there is some uncertainty on the underlying reference order.
The modal reference order is (7,1,2,3,4,6,5), with slightly more than 0.4
posterior probability. However, when we compared the plausibility 
of the observed diagnostic statistic with the reference distribution under the fitted EPL,
we got a warning with a bootstrap classical $p$-value
approximately equal to 0.011. This should indeed cast some doubt on the
use of PL or EPL as a suitable model for the entire dataset. Likely, a more flexible model is to be looked for, such as a mixture of EPL or an alternative parametric class. In fact, we have verified
that, after suitably splitting the dataset into two groups according to
the EPL mixture methodology suggested by \cite{Mollica:Tardella} (best fitting 2-component EPL mixture with BIC=2131.20),
we have a different more comfortable perspective for using
the EPL distribution to separately model the two clusters. The
modal reference orders are
(1,2,3,4,5,6,7) and (1,2,3,7,4,5,6) and the estimated Borda orderings are
(7,6,4,5,3,1,2) and (1,2,3,4,6,7,5), indicating opposite preferences in the two subsamples towards team and individual sports. In this case, no warning by the diagnostic tests applied separately to the two subsamples is obtained, since the resulting $p$-values are
0.991 and 0.677.

\section{Conclusions}
\label{s:conc}

We have addressed some relevant issues in modelling choice behavior and preferences. In particular, the widespread use of the standard Plackett-Luce model for complete  rankings relies on the hypothesis that the probability of a particular ordering does not depend on the subset of items from which one can choose (Luce's Axiom).
This can be considered a very strong assumption and many attempts have been done to develop more flexible models. In particular, \cite{Mollica:Tardella} explored the possibility of adding flexibility with the help of two main ideas: i) the use of a discrete parameter, the reference order, specifying the order of the ranks sequentially assigned by the individual and which should be inferred from the data; ii) the finite mixture of Plackett-Luce distributions enriched with the reference order parameter (EPL mixture). 

In this paper, we have focussed in developing the methodology to infer the reference order of a Plackett-Luce distribution within the Bayesian framework, with an additional monotonicity restriction on the sequence of the alternative choices of the ranks. We have highlighted some difficulties in implementing a well-mixing MCMC approximation, but we have devised appropriately tuned and combined MH kernels. This allows for an easier evaluation of the uncertainty on the EPL parameters, especially on the reference order, which has not been addressed earlier in the literature. Our contribution allows to gain more insights on the sequential mechanism of formation of preferences, whether or not it is appropriate at all and whether it privileges a more or less naturally ordered assignment of the most extreme ranks. In other words, we show how it is possible to assess with a suitable statistical approach the formation of ranking preferences and answer through a statistical model the following questions: ``What do I start ranking firs? The best or the worst? And what do I do then?''. 

We have finally derived a diagnostic tool which can be useful to test the appropriateness of the EPL distribution. Its effectiveness has been checked with applications to both simulated and real examples.

As possible future developments, several directions can be contemplated to further extend the Bayesian EPL with order constraints. First, the methodology can be generalized to accomodate for partial orderings and the introduction of item-specific and individual covariates, that can improve the characterization of the preference behavior. Moreover, a Bayesian EPL mixture could fruitfully support the identification of a cluster structure in the observed sample.

\newpage


\begin{thebibliography}{15}
\newcommand{\enquote}[1]{``#1''}
\providecommand{\natexlab}[1]{#1}
\providecommand{\url}[1]{\texttt{#1}}
\providecommand{\urlprefix}{URL }
\expandafter\ifx\csname urlstyle\endcsname\relax
  \providecommand{\doi}[1]{doi:\discretionary{}{}{}#1}\else
  \providecommand{\doi}{doi:\discretionary{}{}{}\begingroup
  \urlstyle{rm}\Url}\fi
\providecommand{\eprint}[2][]{\url{#2}}

\bibitem[{Alvo and Yu(2014)}]{Alvo}
Alvo M, Yu PL (2014).
\newblock \emph{Statistical methods for ranking data}.
\newblock Springer.

\bibitem[{Caron and Doucet(2012)}]{Caron:Doucet}
Caron F, Doucet A (2012).
\newblock \enquote{Efficient {B}ayesian inference for {G}eneralized
  {B}radley-{T}erry models.}
\newblock \emph{J. Comput. Graph. Statist.}, \textbf{21}(1), 174--196.
\newblock ISSN 1061-8600.
\newblock \doi{10.1080/10618600.2012.638220}.

\bibitem[{Critchlow \emph{et~al.}(1991)Critchlow, Fligner, and
  Verducci}]{Fligner:Verducci-MathPsycho}
Critchlow DE, Fligner MA, Verducci JS (1991).
\newblock \enquote{Probability models on rankings.}
\newblock \emph{Journal of Mathematical Psychology}, \textbf{35}(3), 294--318.

\bibitem[{Diaconis(1988)}]{Diaconis:LectureNotes}
Diaconis PW (1988).
\newblock \emph{Group representations in probability and statistics}, volume~11
  of \emph{IMS Lecture Notes Monogr. Ser.}
\newblock Inst. of Math. Stat.
\newblock ISBN 0-940600-14-5.

\bibitem[{Gormley and Murphy(2006)}]{Gormley:Murphy-Royal}
Gormley IC, Murphy TB (2006).
\newblock \enquote{Analysis of {I}rish third-level college applications data.}
\newblock \emph{Journal of the Royal Statistical Society: Series A},
  \textbf{169}(2), 361--379.
\newblock ISSN 0964-1998.
\newblock \doi{10.1111/j.1467-985X.2006.00412.x}.

\bibitem[{Henery(1981)}]{Henery-Royal}
Henery RJ (1981).
\newblock \enquote{Permutation probabilities as models for horse races.}
\newblock \emph{Journal of the Royal Statistical Society: Series B (Statistical
  Methodology)}, \textbf{43}(1), 86--91.
\newblock ISSN 0035-9246.

\bibitem[{Luce(1959)}]{Luce}
Luce RD (1959).
\newblock \emph{Individual choice behavior: A theoretical analysis}.
\newblock John Wiley \& Sons Inc.

\bibitem[{Marden(1995)}]{Marden}
Marden JI (1995).
\newblock \emph{Analyzing and modeling rank data}, volume~64 of
  \emph{Monographs on Statistics and Applied Probability}.
\newblock Chapman \& Hall.
\newblock ISBN 0-412-99521-2.

\bibitem[{Mollica and Tardella(2014)}]{Mollica:Tardella}
Mollica C, Tardella L (2014).
\newblock \enquote{Epitope profiling via mixture modeling of ranked data.}
\newblock \emph{Statistics in Medicine}, \textbf{33}(21), 3738--3758.
\newblock ISSN 0277-6715.
\newblock \doi{10.1002/sim.6224}.

\bibitem[{Mollica and Tardella(2017)}]{Mollica:Tardella2017}
Mollica C, Tardella L (2017).
\newblock \enquote{{B}ayesian mixture of {P}lackett-{L}uce models for partially
  ranked data.}
\newblock \emph{Psychometrika}, \textbf{82}(2), 442--458.
\newblock ISSN 0033-3123.
\newblock \doi{10.1007/s11336-016-9530-0}.

\bibitem[{Plackett(1975)}]{Plackett}
Plackett RL (1975).
\newblock \enquote{The analysis of permutations.}
\newblock \emph{Journal of the Royal Statistical Society: Series C (Applied
  Statistics)}, \textbf{24}(2), 193--202.
\newblock ISSN 0035-9254.

\bibitem[{Stern(1990)}]{Stern1990}
Stern H (1990).
\newblock \enquote{Models for distributions on permutations.}
\newblock \emph{Journal of the American Statistical Association},
  \textbf{85}(410), 558--564.

\bibitem[{Vigneau \emph{et~al.}(1999)Vigneau, Courcoux, and
  Semenou}]{Vigneau1999}
Vigneau E, Courcoux P, Semenou M (1999).
\newblock \enquote{Analysis of ranked preference data using latent class
  models.}
\newblock \emph{Food quality and preference}, \textbf{10}(3), 201--207.

\bibitem[{Vitelli \emph{et~al.}(2014)Vitelli, S{\o}rensen, Crispino, Frigessi,
  and Arjas}]{Vitelli}
Vitelli V, S{\o}rensen {\O}, Crispino M, Frigessi A, Arjas E (2014).
\newblock \enquote{Probabilistic preference learning with the Mallows rank
  model.}
\newblock \emph{arXiv preprint arXiv:1405.7945}.

\bibitem[{Yu \emph{et~al.}(2005)Yu, Lam, and Lo}]{Yu2005}
Yu PLH, Lam KF, Lo SM (2005).
\newblock \enquote{Factor analysis for ranked data with application to a job
  selection attitude survey.}
\newblock \emph{Journal of the Royal Statistical Society: Series A (Statistics
  in Society)}, \textbf{168}(3), 583--597.

\end{thebibliography}

 \newcommand{\noop}[1]{}


\end{document}